\documentclass[twocolumn,showpacs]{revtex4}

\usepackage{graphicx}
\usepackage{dcolumn}
\usepackage{amsmath}

\makeatletter
\def\btt#1{\texttt{\@backslashchar#1}}%
\DeclareRobustCommand\bblash{\btt{\@backslashchar}}%
\makeatother

\begin{document}

\title[]{Coexistence of antiferromagnetism and superconductivity in heavy-fermions systems}

\author{Y.~Kitaoka$^1$, Y.~Kawasaki$^1$, T.~Mito$^1$, S.~Kawasaki$^1$, G.~-q.~Zheng$^1$, K.~Ishida$^1$, D.~Aoki$^2$, Y.~Haga$^3$, R. Settai$^2$, Y.~\=Onuki$^{2,3}$, C.~Geibel$^4$, F.~Steglich$^4$}

\affiliation{
$^1$Department of Physical Science, Graduate School of Engineering Science, Osaka University, Toyonaka, Osaka 560-8531, Japan\\$^2$Department of Physics, Graduate School of Science, Osaka University, Toyonaka, Osaka 560-0043, Japan\\$^3$Advanced Science Research Center, Japan Atomic Energy Research Institute, Tokai, Ibaraki 319-1195, Japan\\$^4$Max-Planck Institute for Chemical Physics of Solids, D-01187 Dresden, Germany}

\date{\today}

\begin{abstract}
We report the novel pressure($P$)$-$temperature($T$) phase diagrams of antiferromagnetism (AF) and superconductivity (SC) in CeRhIn$_5$, CeIn$_3$ and CeCu$_2$Si$_2$ revealed by the NQR measurement. In the itinerant helical magnet CeRhIn$_5$, we found that the N\'eel temperature $T_N$ is reduced at $P \geq$ 1.23 GPa with an emergent {\it pseudogap} behavior. The coexistence of AF and SC is found in a narrow $P$ range of 1.63 $-$ 1.75 GPa, followed by the onset of SC with line-node gap over a wide $P$ window $2.1 - 5$ GPa. In CeIn$_3$, the localized magnetic character is robust against the application of pressure up to $P \sim$ 1.9 GPa, beyond which the system evolves into an itinerant regime in which the resistive superconducting phase emerges. We discuss the relationship between the phase diagram and the magnetic fluctuations. In CeCu$_2$Si$_2$, the SC and AF coexist on a microscopic level once its lattice parameter is expanded. We remark that the underlying {\it marginal} antiferromagnetic state is due to collective magnetic excitations in the superconducting state in CeCu$_2$Si$_2$. An interplay between AF and SC is discussed on the SO(5) scenario that unifies AF and SC. We suggest that the SC and AF in CeCu$_2$Si$_2$ have a common mechanism.
\end{abstract}


                       
\maketitle
\section{Introduction}
Pressure-induced superconductivity was discovered in the magnetic heavy-fermion (HF) compounds, such as CeCu$_2$Ge$_2$, CePd$_2$Si$_2$, CeIn$_3$, CeRhIn$_5$ and UGe$_2$ \cite{jaccard92,Mathur98,grosche96,hegger00,Saxena01}. There is increasing evidence that the superconductivity in HF compounds takes place nearby a magnetic phase or even coexists with it. By applying a pressure ($P$), a magnetic order is suppressed at a critical pressure $P_c$, where superconductivity is induced in the antiferromagentic (AFM)-HF compounds, such as CeCu$_2$Ge$_2$,\cite{jaccard92} CePd$_2$Si$_2$,\cite{Mathur98}, CeIn$_3$,\cite{grosche96} and CeRhIn$_5$ \cite{hegger00}. In order to gain profound insight into a relationship between magnetism and superconductivity in HF systems, systematic NMR/NQR experiments under $P$ are important, since they can probe the evolution of the magnetic properties toward the onset of SC phase and the character of spin fluctuations in the SC state. 

The superconductivity in underlying compound CeCu$_2$Si$_2$ is considered to occur just at the border of the AFM phase at ambient pressure ($P = 0$) \cite{jaccard92}. This was also corroborated by the existence of the magnetic phase (called as A-phase at zero magnetic field) adjacent to the SC state \cite{Nakamura88,Uemura89}. Furthermore, when magnetic field ($H$) exceeds its upper critical field $H_{\rm c2}$ , the $H$-induced normal state is replaced by magnetic {\it phase A} and successive {\it phase B} with increasing $H$ \cite{Bruls94}.

From the Cu-NQR measurements \cite{ishida99}, we showed that the A-phase at $H$ = 0 behaves as a {\it marginal} AFM state, where slowly fluctuating AFM waves propagate over a long distance and coexist with the SC state. It is quite unusual that their characteristic frequencies range in NQR frequencies of 3 $\sim$ 4 MHz\@. Therefore, this {\it marginal} AFM state, that sets in at $T_m \sim 1.2$ K well above $T_c = 0.65$ K, causes the wipe-out effect in NQR intensity \cite{Nakamura88,ishida99,kawasaki01}. It was remarked that such extremely low-frequency magnetic fluctuations are also reported in high-$T_c$ cuprates La$_{2-x}$Sr$_x$CuO$_4$ where the wipe-out phenomenon of Cu-NQR intensity was also reported \cite{hunt99,curro00}. Thereby, the emergence of slowly fluctuating AFM waves seem to be one of characteristics for strongly correlated electron systems nearby the AFM phase. Recently, this exotic magnetic and SC properties have been extensively investigated through the transport, thermal, and NQR measurements under $P$ \cite{kawasaki01,steglich96,gegenwart98}. At pressures exceeding $P \sim$ 0.2 GPa, the HF-SC state with line-node gap is recovered, coincident with the suppression of the slowly fluctuating AFM waves in Ce$_{0.99}$Cu$_{2.02}$Si$_2$ \cite{kawasaki01}. Its normal state was well described in terms of the self-consistent renormalized AFM spin-fluctuation (SCR) theory \cite{moriya92}.

On the other hand, the Ge substitution for Si expands the lattice of the undoped sample to reduce a strength $g$ of the hybridization between $4f$ electrons and conduction electrons, corresponding to applying a negative chemical pressure\cite{trovarelli97}. Contrary to applying $P$ to increase $g$, it would be expected that magnetic (RKKY) interactions of $4f$ moments become stronger as Ge content increases. In fact, in the previous paper \cite{kitaoka01}, we reported that the 2\% Ge substitution leads to the emergence of AFM order at $T_N = 0.75$ K, followed by the SC transition at $\sim$ 0.4 K\@. The shape of NQR spectrum, that is significantly broadened below $T_N$, revealed that the AFM region  are {\it not spatially separated} from the non-AFM SC one. Thereby, the coexistence of both was shown to occur on a microscopic level. Because of this, the $T$ dependence of $1/T_1$ below $T_c$ does not obey a $T^3$ behavior, demonstrating to differ from the line-node-gap state revealed in the HF-SC regime induced by applying $P$ \cite{kawasaki01}. Persistence of magnetic excitations even in the SC state was thus evidenced from the significant enhancement in nuclear spin lattice relaxation rate $1/T_1$ below $T_c$\@.

In the literature \cite{kitaoka01}, we have proposed that (1) the exotic magnetic SC phases in Ce$_{0.99}$Cu$_{2.02}$Si$_2$ which coexists with the {\it marginal} AFM state, (2) the microscopic coexistence of AFM and SC phase in the slightly Ge substituted compounds, and (3) the $H$-induced  {\it phase A} are accounted for on the basis of the SO(5) theory that unifies superconductivity and antiferromagnetism \cite{kitaoka01,zhang}. We believe that this model could shed light on current ideas regarding the magnetically mediated mechanism of the superconductivity in other strongly correlated electron systems besides the high-$T_c$ cuprates. In establishing whether or not these belong to a new phase of matter, however, further experiments are needed.

Remarkably, a recent $\mu$SR experiment has clearly demonstrated that the singlet superconductivity in Ce$_{0.99}$Cu$_{2.02}$Si$_2$ coexists with the magnetic $A$-phase, or the {\it marginal} AFM state in our terminology, and it exhibits an intimate connection to the magnetism in Ce$_{0.99}$Cu$_{2.02}$Si$_2$ \cite{koda}. This result  has strongly suggested that the superconductivity and the $A$-phase magnetism in CeCu$_2$Si$_2$ has a common background. In other words, we may remark that {\it collective magnetic excitations} inherent to the unconventional SC state exist at the magnetic critical point. 

Recently Hegger {\it $et$ $al$.} found that a new HF material CeRhIn$_5$ consisting of alternating layers of CeIn$_3$ \cite{hegger00} and RhIn$_2$ reveals an AFM-to-SC transition at a  critical pressure $P_c$ = 1.63 GPa lower than in all previous examples \cite{jaccard92,Mathur98,grosche96}. The SC transition temperature $T_c=$ 2.2 K is the highest one to date among $P$-induced superconductors \cite{hegger00}. This finding has opened a way to investigate the $P$-induced evolution of both magnetic and SC properties over a wide $P$ range. The $^{115}$In NQR study of CeRhIn$_5$ has clarified the $P$-induced anomalous magnetism and unconventional superconductivity \cite{mito01}. In the AFM region, the N\'eel temperature $T_N$ exhibits a moderate variation, while the internal field $H_{int}$ at In(1) site in the CeIn$_3$ plane due to the magnetic ordering is linearly reduced in $P = 0 - 1.23$ GPa, extrapolated to zero at $P^* = 1.6\pm 0.1$ GPa. This $P^*$ is comparable to $P_c = 1.63$ GPa at which the SC signature appears, which was indicative of a second-order like  AFM-to-SC transition rather than the first-order one suggested previously \cite{hegger00}. At $P = 2.1$ GPa, it was found that $1/T_1$ reveals a $T^3$ dependence below the SC transition temperature $T_c$, which shows the existence of line-nodes in the gap function \cite{mito01}.

On the other hand, CeIn$_3$ crystallizes in the cubic AuCu$_3$ structure and orders antiferromagnetically ($T_N = 10$ K) at $P = 0$ with an ordering vector {\bf Q}=(1/2,1/2,1/2) \cite{morin88}. $T_N$ monotonically decreases with $P$ and superconductivity appears below $T_c\sim$ 0.2 K at a critical pressure $P_c = 2.55$ GPa, but the onset of the superconductivity was observed {\it only} by the resistivity measurement and is limited in a narrow $P$ range of about 0.5 GPa \cite{Mathur98,grosche96,Walker}. The previous In-NQR result on CeIn$_3$ revealed that the AFM order disappears around $P^*$ = 2.44 GPa close to $P_c$ and the Fermi-liquid behavior was observed below 10 K at $P = 2.74$ GPa \cite{Kohori-2}. However, no SC transition was seen. Thus, the bulk nature of the $P$-induced SC state in CeIn$_3$ has not been confirmed yet by other measurements than resistivity.
This is in contrast with the case of CeRhIn$_5$ where the bulk nature of the SC phase was fully established \cite{mito01} which continues  up to $\sim$5 GPa \cite{Muramatsu}.

The interplay between magnetism and superconductivity is one of the typical and fascinating examples that have revealed the incredible behavior of strongly correlated electrons in condensed matters. The NQR experiments under $P$ is providing fruitful set of data on the evolution from the AFM to SC phase and the character of spin fluctuations in the SC state. In this paper, we compare systematically these $P - T$ phase diagrams in CeRhIn$_5$, CeIn$_3$ and CeCu$_2$Si$_2$ in order to present what is the news.

\section{CeRhIn$_5$}
The $T$ dependence of $1/T_1T$ below 12 K is shown in Fig.1 at $P$ = 0, 0.46, 1 .23 and 2.1 GPa. At $P$ = 0, 0.46 and 1.23 GPa \cite{shinji}, it is clearly seen that an AFM order occurs at 3.8, 4.0 and 3.6 K, respectively, as indicated by a clear peak in $1/T_1T$ due to critical magnetic fluctuations toward the AFM ordering. 
\begin{figure}[h]
\centering
\includegraphics[scale=0.3]{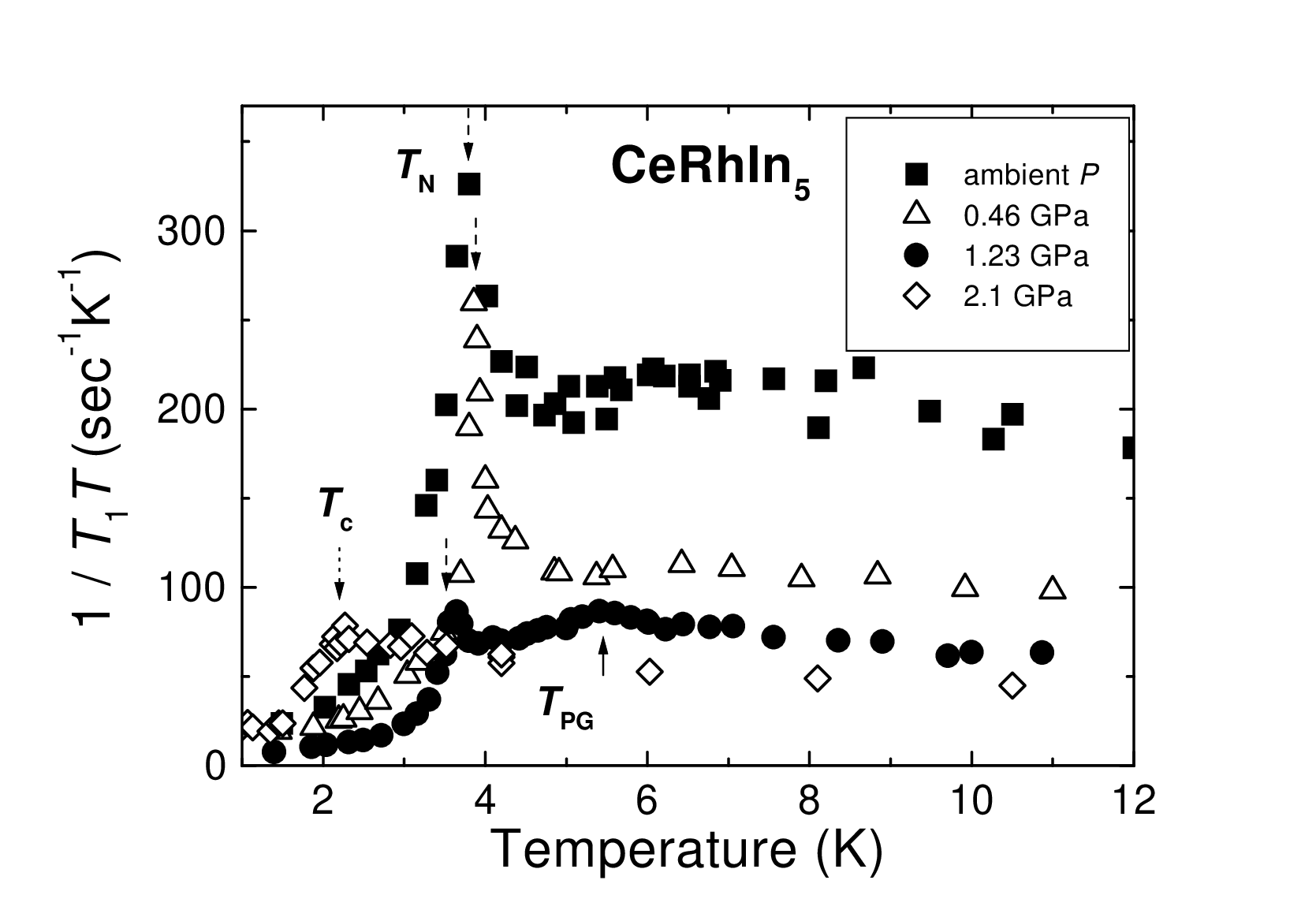}
\caption[]{Temperature dependence of $1/T_1T$ in CeRhIn$_5$.  The solid, broken and dotted arrows indicate $T_{PG}$, $T_N$ and $T_c$, respectively.}
\end{figure}
This indicates that $T_N$ reduces at $P$ = 1.23 GPa as $P$ approaches $P_c$ at which superconductivity sets in. This reduction in $T_N$ is not seen from the previous resistivity measurements \cite{hegger00}. At $P =$ 1.23 GPa,  $1/T_1T$ shows a broad peak around $T_{PG}$ above $T_N$. This resembles the {\it pseudogap} behavior found in the high-$T_c$ copper oxide superconductors \cite{Timusk}.
Namely, when $P$ approaches the critical pressures $P^*$ or $P_c$, the low-energy spectral weight of magnetic fluctuations is suppressed before an ordering occurs. We note that the {\it pseudogap} behavior has been found in either two- or lower-dimensional strongly correlated electron systems \cite{Timusk}.
Very recently, in CeRhIn$_5$, anisotropic three dimensional AFM fluctuation  was reported from neutron scattering at $P$ = 0 with an energy scale of less than 1.7 meV for temperature as high as 3~$T_c$ \cite{Bao-2}. On the other hand, as $P$  further increases up to $P$ = 2.1 GPa where the bulk SC transition appears, $1/T_1T$ continues to increase down to $T_c = 2.2$ K without any signature for the pseudogap behavior as seen in Fig.1. The $T$ variation of $1/T_1T$ is consistent with the three dimensional AFM Fermi-liquid model of the self-consistent renormalized  (SCR) theory for nearly AFM metals \cite{mito01}.
Thus the $P$-induced evolution in the magnetic fluctuations, from a magnetic regime of reduced dimensionality to a more isotropic one, may take place in a narrow $P$ window of $1.2 - 2.1$ GPa, when the AFM order evolves into the SC one.
Recent precise NQR measurement in this $P$ window has revealed  a homogeneous coexistence of AFM and SC order on a microscopic level \cite{mito02}.

\section{CeIn$_3$}
Fig.2 indicates the $T$ dependence of $1/T_1T$ \cite{shinji}. For all values of $P$,  $1/T_1T$ increases down to $T_N$, giving no indication for the pseudogap behavior.
\begin{figure}[h]
\centering
\includegraphics[scale=0.3]{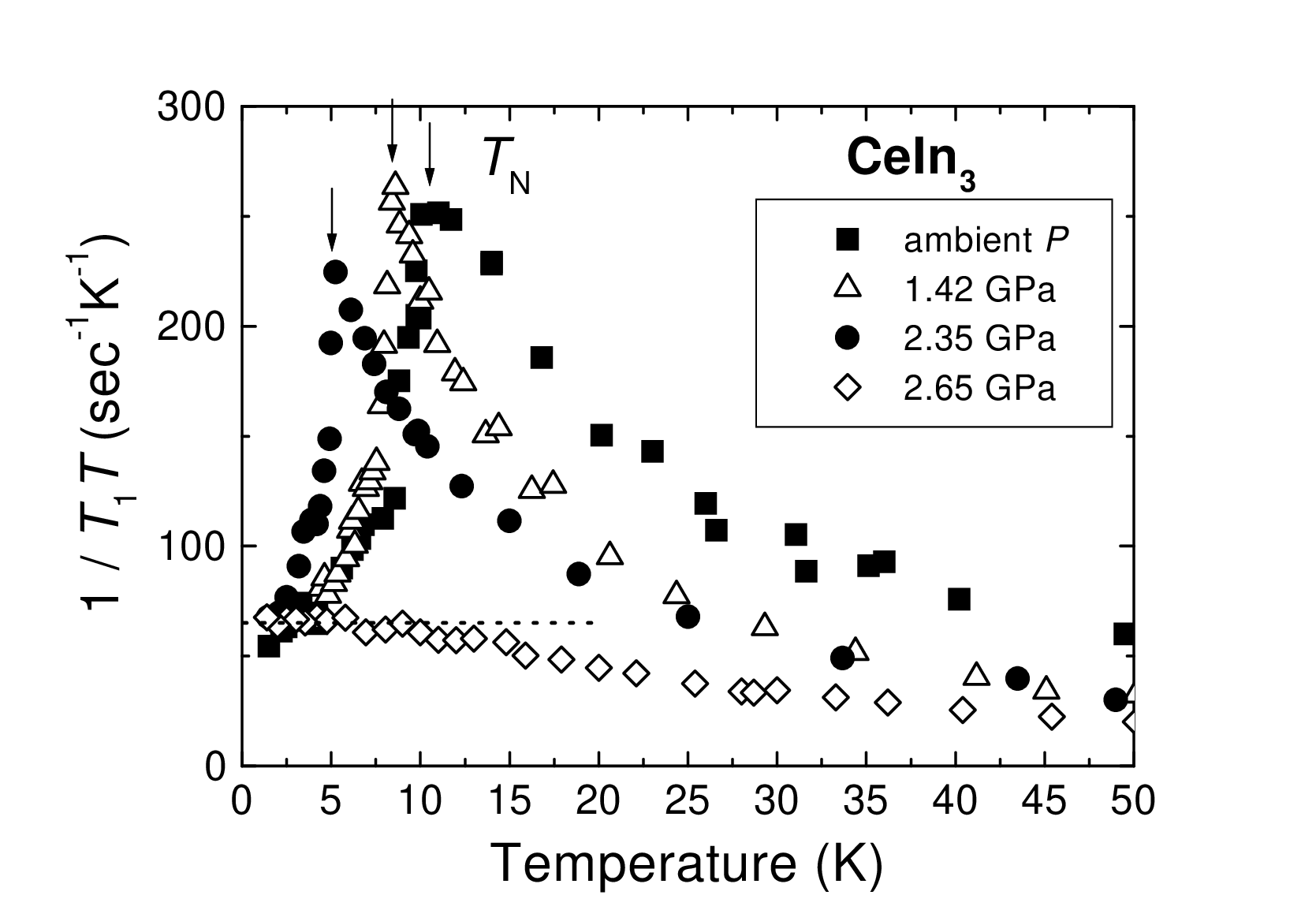}
\caption[]{Temperature dependence of $1/T_1T$ in CeIn$_3$. The solid  arrows indicate $T_N$. Dotted line is eye guide.}
\end{figure}
At $P = 2.35$ GPa where the zero resistance is observed, it has  been confirmed that the magnetically ordering  survives with a relatively high value of 
$T_N = 5$ K. The SC transition is, however, not found by the high-frequency ac-susceptibility measurement down to 100 mK using the {\it in-situ} NQR coil, despite that a zero resistance was observed in the same sample at $T_c\sim$ 0.15 K.
From the present experiment, we conclude that the $P - T$  phase 
diagram and the nature of the SC phase in CeIn$_3$ differs from in CeRhIn$_5$, in many aspects, reflecting 
their contrasting electronic and magnetic properties; we speculate that the SC phase in CeIn$_3$ that accompanies the Meissner diamagnetism, if any, is even narrower than suggested by the resistivity measurement \cite{shinji}.

\section{The AF- and SC-phase diagrams of CeRhIn$_5$ and CeIn$_5$}
The respective $P-T$ phase diagrams for CeRhIn$_5$ and CeIn$_3$ are presented in Figs.3a and 3b \cite{shinji}.  For CeRhIn$_5$, $T^*$, that is scaled to the bandwidth of HF state,  slightly increases with increasing $P$ up to 1.0 GPa, as does $T_{N}$ which coincides with the previous result \cite{hegger00}.
However, $T_N$ decreases above $P = 1.23$ GPa. At the same time, a {\it pseudogap} behavior emerges below $T_{PG} \sim 5.5$ K.
\begin{figure}[h]
\centering
\includegraphics[scale=0.42]{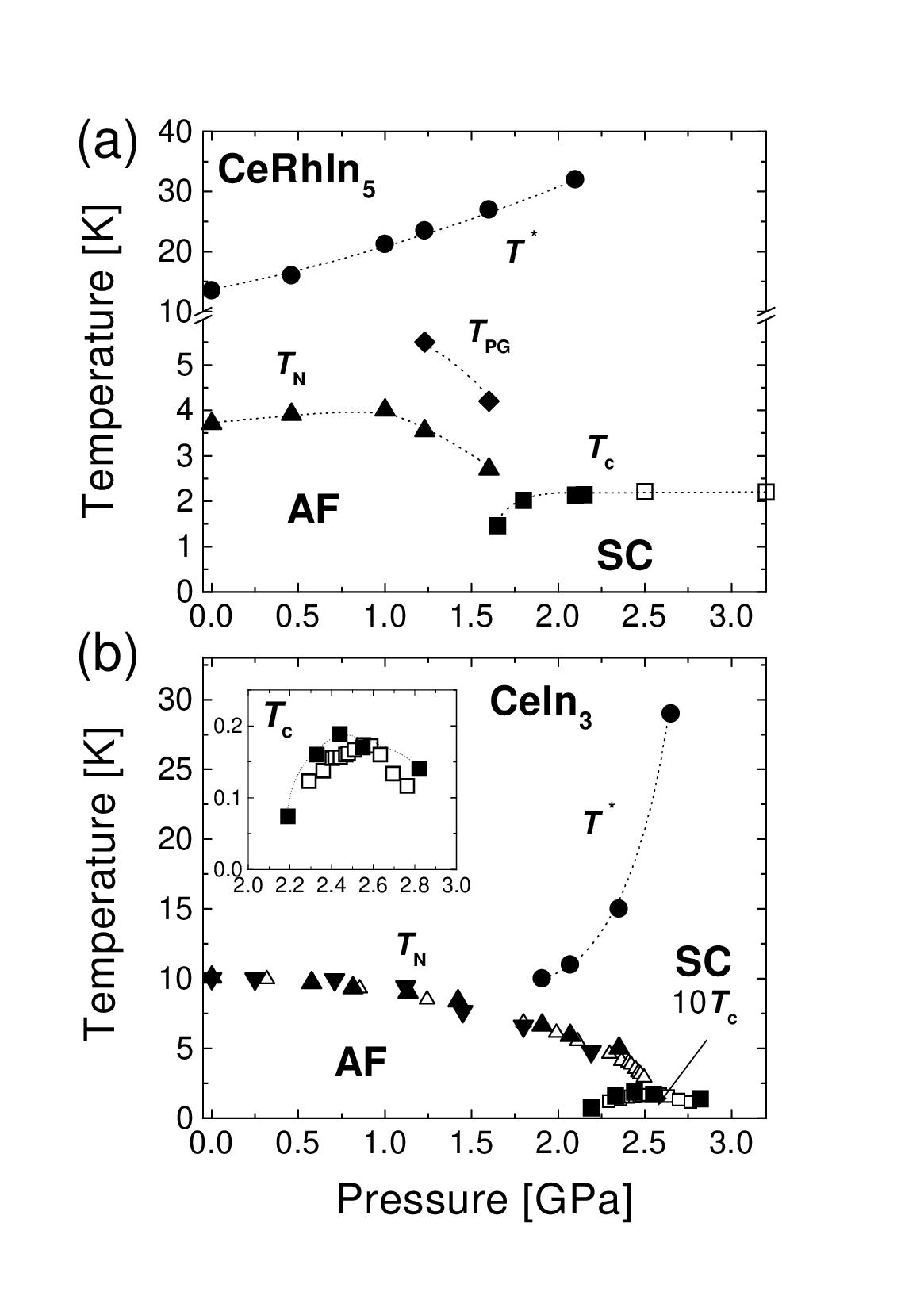}
\caption[]{Pressure$-$temperature phase diagrams (a) for CeRhIn$_5$ and (b) for CeIn$_3$.
(a) The open marks are determined from the resistivity measurements \cite{Muramatsu}, and the solid squares are determined from the ac-$\chi$ measurement \cite{mito01}.
(b) The open marks for $T_N$ and $T_c$ are taken from the resistivity measurements \cite{Mathur98,grosche96,Walker}.
The rest marks  are determined from the present work.
The inset indicates the detailed $P$ dependence of $T_c$ in expanded scales.
Dotted lines in both figures are eye guides.}
\end{figure}
As $P$ is further increased, $T^*$ moderately increases with $d T^*/d P \sim $8K/GPa. It is noted that $T^*$ is equivalent to the temperature $T_{max,\rho}$ at which the resistivity exhibits a maximum value and also $dT^*/d P \sim dT_{max,\rho}/dP$ \cite{hegger00}.
We remark that these values are close to those in CeCu$_2$Si$_2$ \cite{kawasaki01} which is a superconductor at $P = 0$ and reveals a SC state over a wide $P$ region as well \cite{Thomas}.
In CeRhIn$_5$, the anisotropic 3D-AFM fluctuations may survive until the system is close to the SC state, where the {\it pseudogap} behavior is emergent. 
Above $P^*$ or $P_c$ where the bulk superconductivity appears, the AFM spin correlations  become more isotropic and the $T$ dependence of 1/$T_1T$ for $T>T_c$ can be explained on the basis of the 3D-SCR theory for nearly AFM metals \cite{mito01}.
While the isotropic AFM fluctuation regime is fully established, the bulk SC is insensitive against $P$.
For CeIn$_3$, the steep increase of $T^*$ above $P$ = 1.9 GPa indicates an evolution of the system into an itinerant magnetic regime where the SC state becomes emergent.
Moreover, close to $P_c$, the normal-state resistivity, $\rho$ in CeIn$_3$ at low $T$ is also consistent with the 3D-AFM fluctuation model \cite{grosche96}. However, when the pressure is further increased above $P_c$, a Fermi-liquid behavior of the resistivity returns more rapidly in CeIn$_3$ than in CeRhIn$_5$ and CeCu$_2$Si$_2$ \cite{grosche96}.
This is also corroborated by the observation of $T_1T$ = constant behavior at $P = 2.65$ GPa in CeIn$_3$.
We should remark that an upper temperature scale $T_{max,\rho}$ below which local magnetic fluctuations are increasingly frozen out, is as high as $70 - 100$ K over the $P$ range from 0 to 3 GPa \cite{grosche96}.
This large difference between $T^*$ and $T_{max,\rho}$ in CeIn$_3$ contrasts strongly with the coincidence of the both in CeRhIn$_5$, resulting in the steeper change of $T^*$ as a function of $P$ in CeIn$_3$ compared with those in CeRhIn$_5$ and CeCu$_2$Si$_2$.
A reason for such a narrow SC region in CeIn$_3$ is proposed as due to the small window for the 3D-AFM fluctuation regime because of its large rate of $dT^*/dP$. This small window for the 3D-AFM fluctuation regime  against $P$ in CeIn$_3$ may be related to its cubic crystal structure which is more sensitive to external $P$ than a tetragonal structure. 

Based on magnetically mediated SC theoretical models \cite{Model}, it is predicted that 2D-AFM fluctuations are superior to 3D-AFM fluctuations in producing SC  \cite{Monthoux}. Therefore, the enhancement of $T_c$ in layered CeMIn$_5$ over CeIn$_3$ has been suggested to be due to their quasi-2D structure \cite{hegger00,Cornelius,Petrovic}. Here we propose that the small window for the spin fluctuations regime in CeIn$_3$ may, at least partly,  be responsible for its  lower $T_c$ , which should also be taken into account in modeling the SC in these compounds.
\section{The AF and SC phase diagram in CeCu$_2$Si$_2$}
Fig.4 shows the phase diagram determined as functions of Ge content $x$ for CeCu$_2$(Si$_{1-x}$Ge$_x$)$_2$ and of $P$ at $x$ = 0\@ \cite{kawasaki02}.
Here we expect that a primary effect of the Ge doping is to expand the lattice and that its chemical pressure is $-7.6$ GPa per 100\% Ge doping as suggested from the $P$ dependence of $\nu_Q$ in CeCu$_2$Ge$_2$ and CeCu$_2$Si$_2$.\cite{kitaoka95}
\begin{figure}[h]
\centering
\includegraphics[scale=0.42]{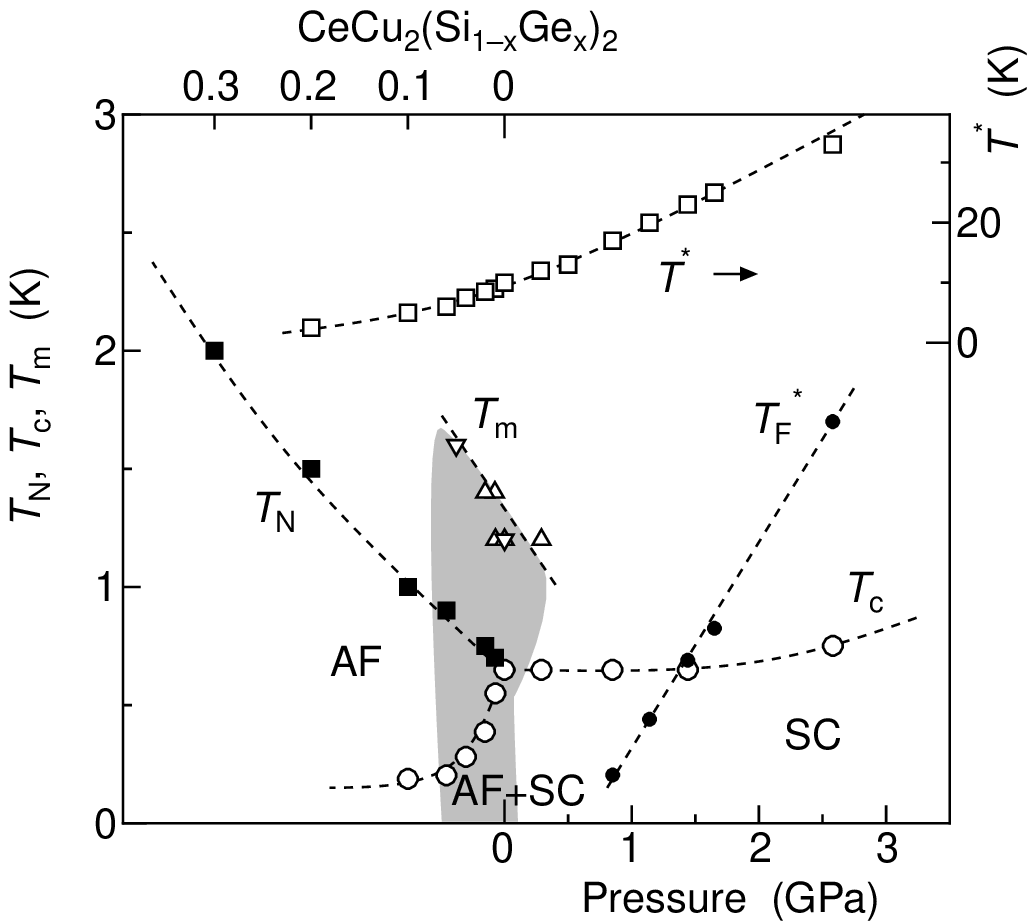}
\caption[]{\protect The AF and SC combined phase diagram for CeCu$_2$(Si$_{1-x}$Ge$_{x}$)$_2$ and for Ce0.99 ($x = 0$) under pressure.
$T_{\rm N}$ (closed squares) and $T_{\rm c}$ (open circles) are the AFM and SC transition temperature, respectively.
Also shown are the characteristic temperature to be scaled to the band width of HF state $T^*$ (open squares) and $T_m$ below which the marginal AFM state is dominant (open triangles).
The data for Ce0.99 under pressure are cited from the previous paper.\cite{kawasaki01}}
\end{figure}
In fact, the recent NQR experiment under $P$ has been carried out on  $x$ = 0.02 where the AFM order below $T_N \sim$ 0.75 K coexists with the superconductivity below $T_c \sim$ 0.4 K\@ \cite{kawasaki02}. As indicated in Fig.5, at pressures exceeding $P$ = 0.56 GPa, the AFM order is suppressed and concomitantly the unconventional SC state evolves into a typical HF one with the line-node gap.
This result demonstrates that the occurrence of the AFM order is due to the lattice expansion caused by doping Ge\@ and that there exist gapless magnetic excitations in the underlying coexisting state of AFM and SC order at $P$ = 0.

The characteristic temperature, $T^*$ which is scaled to the bandwidth of HF state, increases continuously with decreasing $x$ or applying $P$.
$T_N$ is determined as the temperature below which $H_{int}(T)$ appears and at which $1/T_1$ exhibits a peak \cite{kitaoka01,kawasaki02}. 
The AFM order suddenly sets in once $x>0$, whereas, at $x = 0$, it becomes marginal without any trace of $H_{int}$ down to 0.012 K. The marginal AFM state develops below $T_m$ in the ranges of $0\leq x < 0.06$ and $0 < P < 0.2$ GPa, which are shaded by a color in Fig.4. The appearance of it is shown to be limited to the border of the magnetic critical point at $x\sim$ 0. Approaching $x$ = 0, it is noteworthy that the AFM order seems to disappear as a result of $T_N$ approaching $T_c$. Nevertheless, the SC phase can coexist with either the marginal or the  AFM state in $0\leq x < 0.06$, characterized by gapless  magnetic excitations.
This fact indicates that both magnetism and superconductivity has a common background.

In the $P$ range $0 < P < 0.2$ GPa, however, it should be stressed that such the marginal AFM state is expelled below $T_c$ at $H = 0$, but when $H$ 
turns on to suppress the superconductivity, a first-order like SC to magnetic 
{\it phase A} is induced \cite{Bruls94,steglich96,gegenwart98}.   With further increasing $P$ in $P > 0.2$ GPa, the AFM marginal state is completely suppressed and at the same time, the HF superconductivity is recovered, exhibiting the line-node gap. 
 In the shaded zone, the exotic interplay occurs between antiferromagnetism and superconductivity.

Next, we shed further light on the coexisting state by comparing with the typical AFM-HF-SC compound UPd$_2$Al$_3$ where the AFM order sets in below $T_N \sim$ 14.5 K, followed by the onset of typical HF-SC transition below $T_c \sim$ 1.98 K \cite{geibel,sato,metoki}.
In UPd$_2$Al$_3$, the coexistence of the AFM order with the superconductivity has been established by various measurements \cite{sato,tou,matsuda,caspary}.
The Al- and Pd-NQR results showed that $1/T_1$ decreases rapidly due to an opening gap at a part of the Fermi surface below $T_N$ and follows the $1/T_1 \propto T^3$ below $T_c$, consistent with the line-node gap.\cite{tou,matsuda} 
In UPd$_2$Al$_3$, note that an average occupation of $5f$ electrons per U atom is slightly less than three and larger than one in Ce HF systems \cite{yotsuhashi}.
With some relevance with this, nearly two localized $5f$ electrons are responsible for the AFM order with a relatively large spontaneous moment of 0.85$\mu_B$ per U atom.
Since the spin-wave excitations in the AFM state is gapped, the particle-hole excitations in HF state are dominant well below $T_N$, as actually evidenced from the $T_1T = $constant behavior typical for the Fermi-liquid state.
Therefore, the $1/T_1$ in the SC state is not affected by the magnetic excitations even in the coexisting state, but reflects its line-node SC gap structure.
However, the situation in underlying CeCu$_2$Si$_2$ is completely different, because one $4f$ electron per Ce ion play vital roles for both the magnetism and the superconductivity, leading to the novel interplay between the magnetism and the superconductivity revealed in CeCu$_2$Si$_2$.

Finally, we compare the present results with those on Ce$_{0.975}$Cu$_2$Si$_2$ (Ce0.975) where an AFM order was reported below $T_N \sim$ 0.6 K \cite{ishida99,modler95}. Unlike the Ge substituted samples, however, any signature of superconductivity was not observed up to $P$ = 2 GPa in Ce0.975, and the AFM order persists \cite{kawasaki03}.
\begin{figure}[h]
\centering
\includegraphics[scale=0.4]{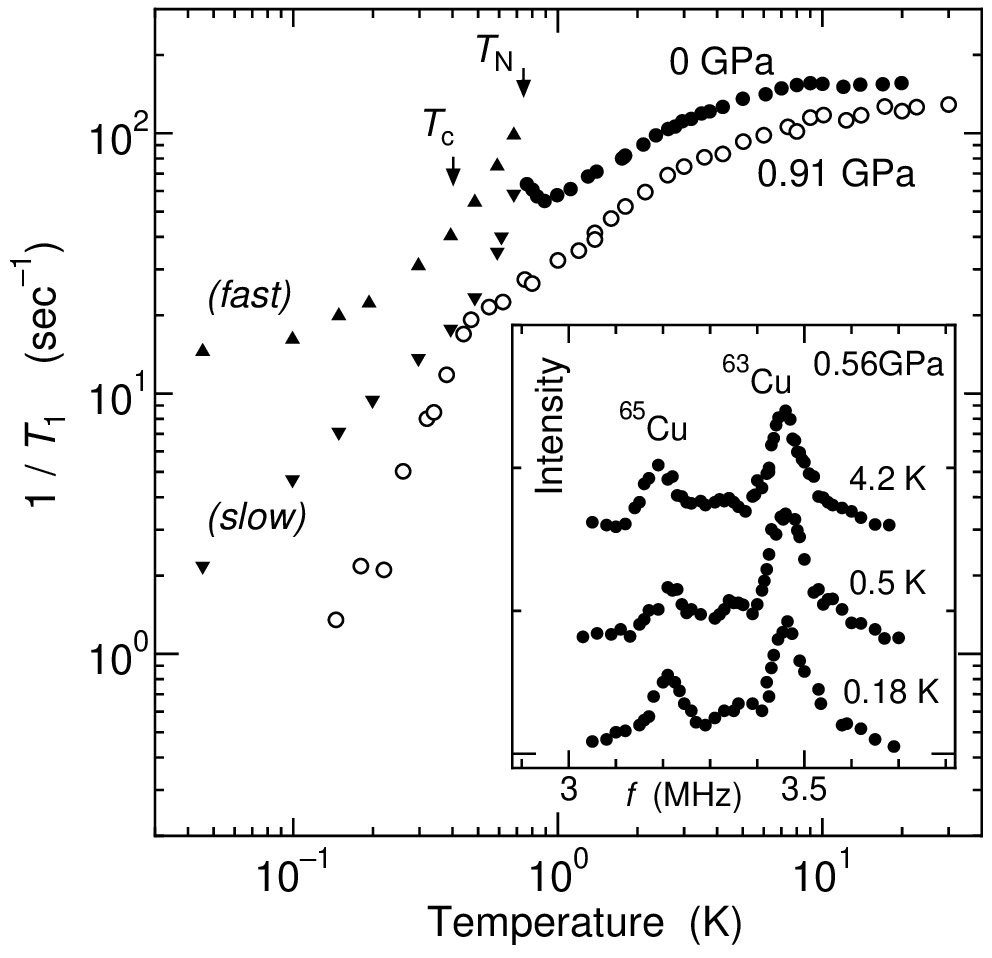}
\caption{Temperature dependence of $1/T_1$ at 0 and 0.91 GPa at $x$ = 0.02 in CeCu$_2$(Si$_{1-x}$Ge$_x$)$_2$\@.
Since $1/T_1$ is not by single component at 0 GPa below $T_N$, both the fast and slow relaxation components are plotted.
The inset shows NQR spectra at 0.56 GPa, indicative of no hyperfine broadening due to an onset of AFM ordering\@.} 
\end{figure}
This is in strong contrast with that the superconductivity is robust against increasing Ge content. Furthermore, the recent NQR result under $P$ at $x = 0.02$ has revealed that the AFM order is suppressed and the HF superconductivity with the line-node gap is recovered at $P$ = 0.56 GPa as seen in Fig.5 \cite{kawasaki02}.
These results indicate that the AFM order in Ce0.975 is driven by some disorder effect that destroys the superconductivity\@.
By contrast, the AFM order at $x = 0.02$ is primarily due to the lattice expansion\@. Therefore, the magnetic order can coexist with the superconductivity without destroying it.

The AFM and SC phases that emerge in the shaded zone including in $0 < P < 0.2$ GPa in Fig.4 may be new states of matter.  
In the literature \cite{kitaoka01}, in fact, we have proposed that the microscopic coexistence of AFM and SC phase in the slightly Ge substituted compounds, and the magnetic field-induced  {\it phase A} that was revealed in the $P$ range $0 < P < 0.2$ GPa are accounted for on the basis of the SO(5) theory that unifies superconductivity and antiferromagnetism.\cite{kitaoka01,zhang}
The present results have revealed, we believe, that the magnetic critical point is located near $x\sim$ 0 and the {\it marginal} AFM state existing in $0 \leq x < 0.06$  below $T_c$ may be identified as a collective magnetic mode in the 
coexisting phase, whereas it in $0 < P < 0.2$ GPa turns out to be {\it competitive} with the onset of SC phase at $H$ = 0 \cite{Bruls94,kawasaki01,steglich96,gegenwart98}.
This latter fact is relevant to the emergence of a $H$-induced {\it first-order} like SC to magnetic {\it phase A} transition \cite{Bruls94,steglich96,gegenwart98}.
Concerning the interplay between magnetism and superconductivity, we would propose that the marginal AFM state in the SC state at $x$ = 0 may correspond to a {\it pseudo Goldstone mode} due to the broken SO(5) symmetry, and to the gapless magnetic excitations due to the broken spin symmetry in the coexisting state in the range of $0 < x < 0.06$.
Due to the closeness to the magnetic critical point, the marginal AFM state in the SC state should be such a gapped mode that is characterized by extremely tiny excitation energy. We thus believe that this SO(5) model could shed light on the intimate interplay between magnetism and superconductivity in CeCu$_2$Si$_2$ that is the long-standing issue over a decade. 
In order to get further insight into the marginal AFM state and the sudden emergence of AFM order that both coexist with the superconductivity, extensive NQR experiments for Ge-doped system under pressure are now in progress.

\section{conclusion}

In CeRhIn$_5$ that is already in the itinerant magnetic regime at $P = 0$, $T_N$ slightly increases with $P$ at lower pressures which is in accordance with the previous report \cite{hegger00}. However, $T_N$ starts to decrease
above $P$ = 1.23 GPa approaching the critical value at which superconductivity sets in. At the same time, the {\it pseudogap} behavior emerges.
By contrast, in CeIn$_3$, the localized magnetic character is robust against the application of $P$ up to 1.9 GPa, beyond which $T^{*}$, which marks  an evolution into an itinerant magnetic regime, increases rapidly. We remark that the SC emerges in such the itinerant regime. The window for the 3D-AF fluctuation regime in CeIn$_3$ is much narrower than in CeRhIn$_5$.
These results suggest that the different robustness of the spin fluctuations regime against $P$ should be taken into account in explaining the difference of $T_c$ in  CeRhIn$_5$ and CeIn$_3$.  

In CeCu$_2$Si$_2$, once its lattice parameter of $x$ = 0 is slightly expanded, the AFM order sets in at $T_N\sim 0.7$ K, followed by the onset of superconductivity below $T_c =$ 0.5 K. This sudden emergence of AFM order due to the slight 1\%-Ge substitution reinforces the existence of the {\it marginal} AFM state in the undoped Ce$_{0.99}$Cu$_{2.02}$Si$_2$ ($x$ = 0) where slowly fluctuating AFM waves propagate over a long distance. The measurement of the NQR spectrum  below $T_N$ is a most reliable way to exclude the presence of phase segregation between a paramagnetic SC and an AFM region. It is remarkable that when $T_N$ approaches $T_c$ at the magnetic critical point ($x\sim 0$), the AFM order suddenly disappears, indicative of a {\it first-order} like transition in the magnetic phase, in contrast with the robustness of the SC one.

The $1/T_1$ result, that does not show any significant reduction below $T_c$, has demonstrated that the SC state coexisting with the AFM order differs from that with the line-node gap.@It reveals that the magnetic excitations are gapless due to the coexistence of AFM and SC phase.
Apparently, the fact that both the marginal AFM state and the AFM order can continue to coexist with the superconductivity strongly suggests that both the magnetism and the superconductivity has a common mechanism.

\section*{Acknowledgements}
We thank A.~Koda, W.~Higemoto, and R.~Kadono for valuable discussions on $\mu$SR results. This work was supported by the COE Research (10CE2004) in Grant-in-Aid for Scientific Research from the Ministry of Education, Culture, Sports, Science and Technology of Japan.
One of the authors (Y.~Kawasaki) has been supported by JSPS Research Fellowships for Young Scientists\@.

\end{document}